# WHY IS THE REFRACTIVE INDEX CANNOT BE NEGATIVE

M.V. Davidovich

*Saratov State University, 410012, Saratov, Russia,*

E-mail *DavidovichMV@info.sgu.ru*

*Abstract* − It has been shown that for left-handed metamaterials and generally for negative refraction media the refraction index cannot be entered unequivocally and cannot be considered as real, and especially as negative. This index for above referred media is not expedient.

The refractive index (RI) $n$ (or retardation coefficient) was initially introduced in optics long before Maxwell has formulated the electrodynamics which interprets the optics as its own part. So, it has been transferred from the scalar optic problems to vector electromagnetic ones. At that time the dispersion was not considered. Recently the so-called left-handed media (LHM) or metamaterials with negative refractions (NR) are under the intensive investigation. The widespread opinion dominates in literature that LHMs have the negative refraction index (NRI). In 1967 V.G. Veselago has published the paper [1], where he considered the medium (which he called left) with scalar real and simultaneously negative permittivity $\varepsilon$ and permeability $\mu$. He investigated the geometric (ray) diffraction theory for infinite in two directions (*x,y*) and finite in *z*-direction plate of thickness *d* with such LHM and has discovered the anomalous refraction Snell law. Also he has discovered some anomalous effects: Doppler effect, Vavilov-Cherenkov effect, negative light pressure. These effects are connected with NR phenomena (excepting negative pressure). It was well-known long before the Veselago's paper and considered in several publications (see references in the papers [2–9]). These considerations proceed from earlier Lamb (1904), Laue (1905), Mandelstam (1940) and others, from the papers corresponding with backward wave tubes and antennas. The history of this question one may find in Russian [2–9] and English [9] papers. The doubtless Veselago's merit is that he drew attention of scientific community to necessity of search possible artificial media (AM) with such unusual properties. Since 80-th to 90-th the research direction of AM investigations is expansive developing. The researches have been begun early in 40 and 50th years and then was named as investigation of artificial dielectrics [9–12]. Next the more general name "metamaterials" was assigned to these AM later. Next the periodic AM with different forms of wire inclusions (wire media) began to be studied intensively in the beginning of 90th and were then are named as metallic photonic crystals (PC).

In 2000 J. Pendry has published the article [13] in which he claimed that the Veselago lens (further in literature named as ideal Pendry lens) is overcoming the diffraction limit. But the Pendry's consideration was based on gross errors (see, for example, the articles [3–6, 14–22] and the discussion there). After the publication [13] and similar, including the experimental work [23], the such conceptions as Veselago medium, double negative materials (DNM, DNG), backward media, LHM, wire PCs, complex media, negative group velocity media (NGV), negative refraction index media and some others finally have been approved. And the number of publications on these questions increases avalanche-like. Although the term "negative refraction" is the most general and it was well-known long before the paper [1], and this phenomenon takes place also in slow wave systems, nature crystals, dielectric PC, generally in optics when the energy transfer direction of monochromatic wave may constitute the obtuse angle with the direction of phase motion, the term "negative refraction index" on our opinion is not correct. We will show further why the RI couldn't be negative, can not be such, and why it is not appropriate for NR media. Authors of some works (apparently, realising it) instead of NRI use the terms "negative refraction" [6] or "negative media" [8] along with NR. The big number of above abbreviations also indicates on the problem. In several such papers the NRI is not considered altogether, but in most of publications the NRI $n<0$ nevertheless is considered. Thus, the NRI is the sufficiently established term, and the number of papers with its usage is highly large. The goal of this paper is to show that the question here is not only in terminology: the introduction of NRI $n<0$ in the relations, in which it evidently couldn't be introduced, often leads to incorrect physical results.

There is prevailing common opinion in literature that for LHM with $\varepsilon<0$ and $\mu<0$ one must extract the root as follow: $n=-\sqrt{\varepsilon\mu}<0$. And in the normalized impedance it is need to take the root branch as $\rho=\sqrt{\mu/\varepsilon}>0$ [24]. These values one can insert in spectral form of Maxwell equations for harmonic plan wave (the equations (5) from [1]), that is indirectly assumed under such determination [24]. The energy flow and phase motion directions then are opposite, i.e. the wave is backward. The values $n$ and $\rho$ are introduced in such manner in optics, But they there both positive (also together with positive $\varepsilon$ and $\mu$). To choose the branch of root one must set some physical condition. In optics there is the dissipation according to which $\text{Im}(\tilde{n})<0$ and $\text{Re}(\tilde{\rho})>0$ (for complex values with the time dependence $\exp(i\omega t)$). It may seem that by introducing the complex RI, impedance, permittivity $\tilde{\varepsilon}=-\varepsilon-i\varepsilon''$ and permeability $\tilde{\mu}=-\mu-i\mu''$, where all quantities are positive, one can from the conditions $\text{Im}(\tilde{n}) < 0$ and $\text{Re}(\tilde{\rho}) > 0$ get the unambiguously: $\tilde{n} = n' - in''$, $n' = -\sqrt{\varepsilon\mu} < 0$, $n'' = (\varepsilon''\mu + \varepsilon\mu'')/\sqrt{\varepsilon\mu} > 0$, $\tilde{\rho} = \sqrt{\mu/\varepsilon}[1 + i(\mu''/\mu - \varepsilon''/\varepsilon)/2]$. But the problem here that it is impossible to take the limits $\varepsilon'' \to 0$ and $\mu'' \to 0$. It will be shown further. Furthermore, the value $\sqrt{\varepsilon\mu}$ is polysemantic. Imposing these estimates and binding two roots (either of the two is double-valued), we use only backward wave. But in the finite PC both backward and forward waves are possible for each dispersion branch. Let's consider the simplest infinite case of propagation along $z$-axis. For such wave the transition from forward to backward ones occurs under the replacement $k_z \to -k_z \pm 2m\pi/a_z$ ($a_z$ is the period along $z$), and also by going over the passage from one dispersion branch (hypersurface) to another through any bandgap by changing the $k_0$. These propagation branches (bands) are separated by the bandgaps, and the waves in different directions are differed (for anisotropic or bianisotropic AM). The effective permittivity and permeability (and RI) are the even functions of $k_z$ (and others components if any). Hence, the transfer from forward wave to backward one takes place not due to sign of $n$, but owing to sign of $k_z$ (under $|k_z a_z| < \pi$), or due to sign of $dk_0/dk_z$ (here we propose the absence of loss). In low frequency limit $k_0 \to 0$ and for $|\mathbf{k}| \to 0$ (in our case for $k_z > 0$) the wave is forward-directed, i.e. the NR corresponds with the characteristic Bragg spatial $|\mathbf{k}| \cdot |\mathbf{a}| \sim 1$ ($\mathbf{a}$ is the translation vector) and frequency $k_0 n_{ef}(0)|\mathbf{a}| \sim 1$ resonances (scales). Here $n_{ef}(0)$ is the effective RI in the low frequency limit which is produced by the homogenization. The exclusion here is the physically unrealizable and similar ideal plasma wire medium with infinite parallel wires, which has the low-frequency cutoff. Accordingly all the RI $n_{ef}$, $\varepsilon_{ef}$ and $\mu_{ef}$ (without dependence from that tensors they or scalars) are depending of $\omega$ and $\mathbf{k}$. Usually the homogenization in optics (excepting crystal optics) leads to the isotropic RI, as the wavelength is sufficiently greater than the typical dimensions of natural substance. In the hard ultraviolet and in the X-ray diapasons it is not so.

It is already well known from optics that in the regions with strong anomalous dispersion it may be negative i.e. the inverse waves may exist. In particular, under the large oscillator forces the NR in narrow band may exist (that runs up to this seldom), but $n'(\omega) = \text{Re}(n(\omega)) > 0$ (see formula 83.13 from [25]). In this case the phase and the energy move in opposite directions, and the losses are highly large, i.e. $n''(\omega) = -\text{Im}(\tilde{n}(\omega))$ may be of $n'(\omega)$ order. The group velocity does not characterize here anything. Usually in periodic waveguides (say, for microwaves) the losses are negligible small, therefore the NR is described by positive retardation $n$ and negative group velocity [26]. These are the one-dimensional-periodic (1-D-P) structures, therefore $n$ is scalar. The losses lead to negative influence on NR media properties (in particular, they destroy the focusing capability of Pendry lens (PL) and Veselago-Pendry lens (VPL)), and the sufficiently big papers are devoted to that. But, as it is surprising, some losses are necessary are necessary for existence of NR. Namely, in the paper [27] it has been shown that in isotropic media with NR there is lower limit of electric and magnetic losses, and the NR does not exist lower this limit. The Kramers-Kronig relations [25] for $n^2(\omega)$ have been used to proof this and the criterion has been obtained [27]:

$$\frac{2}{\pi}\int_0^\infty \frac{\varepsilon''(\tilde{\omega})\mu'(\tilde{\omega})+\mu''(\tilde{\omega})\varepsilon'(\tilde{\omega})}{(\tilde{\omega}^2-\omega^2)}\tilde{\omega}^3 d\tilde{\omega} \leq -1.$$

At that the value $n^2(\omega)$ (but not $n(\omega)$) is the analytical functions in one of $\omega$ semiplanes (depending on the sign choice in the $\exp(\pm i\omega t)$). Nevertheless, to reduce the losses the several metamaterials different from DNG wire media are investigated in a number of papers [28–30]. In these papers the 1-D-P PCs with high-temperature superconductive and magnetic films [28], with superconductive and dielectric films [29], and also with additional inclusions of structures like "magnetic atom" in form of $MgF_2$ film with golden plate [30] have been considered. Such superconductive magnetic structures are the uniaxial PCs. Thus, in [29] the transverse $\varepsilon_\parallel$ and longitudinal $\varepsilon_\perp$ components of permittivity are introduced, and it is shown that even for superconductive state there are sufficiently considerable losses. They suppress the increase of damping (evanescent) mode amplitudes and put obstacles for superresolution, but, nevertheless, the NRI is introduced in [29] and the references to experiments concerning the figure of merit (FOM) in form $|n'|/|n''|$ are given. This FOM for LHM in infra-red and optical diapasons lies in the region $0.1-3.5$. But such anisotropic or bianisotropic structures couldn't be described by one scalar RI. Moreover, the magnetic inclusions need the magnetic field which is proposed to control their properties [28]. Such PCs in magnetic field are the gyrotropic media.

It is useful to remember how the $n$ is introduced in optics. For transparent isotropic media in the disregard of dispersion (and therefore, neglecting of losses) we have $\varepsilon(\mathbf{r})>1$, and it is possible to determine the RI $n=\sqrt{\varepsilon}$. In this case it is a simple constant for homogeneous medium. The taking into account of frequency dispersion for monochromatic waves already leads to complex frequency depended values $\varepsilon(\mathbf{r},\omega)$ и $n(\mathbf{r},\omega) = n'(\mathbf{r},\omega) - jn''(\mathbf{r},\omega)$, where $\varepsilon''(\mathbf{r},\omega) \geq 0$, $n''(\mathbf{r},\omega) \geq 0$, in which connection the equality is possible only if $\omega = 0$ or $\omega \to \infty$ [25]. That is justly in any dissipative medium, and the complex number can't be negative. In his paper [1] V.G. Veselago at first proceeds from the dispersion equation (DE) for anisotropic medium without dissipation [1]:

$$\det A = 0, \quad A_{ik} = k_0^2 \hat{\varepsilon}_{il} \hat{\mu}_{lk} - k^2 \delta_{ik} + k_i k_k, \qquad (1)$$

where $k_0^2 = \omega^2/c^2$ and $k^2 = \mathbf{k}^2$. Besides the equation (1) one may, as a matter of fact, use also the equation

$$\det B = 0, \quad B_{ik} = k_0^2 \hat{\mu}_{il} \hat{\varepsilon}_{lk} - k^2 \delta_{ik} + k_i k_k, \qquad (2)$$

i.e. the DE and $n$ are ambiguously determined and introduced. Further in proposal of isotropy the equation (1) is rewrote in [1] as

$$k^2 - k_0^2 n^2 = 0, \quad n^2 = \varepsilon\mu. \qquad (3)$$

That, as a matter of fact, means the scalarization of Maxwell equations, that generally it is not necessary to do and anyone shouldn't to do, because of only the values $\hat{\varepsilon}$ and $\hat{\mu}$ are initially in these equations (or in the more complicated material equations which must be used there). The DE (1) and (2) are the equations to determine the dispersion, i.e. the dependence $\mathbf{k} = \mathbf{k}(k_0)$ or inverse dependence $k_0 = k_0(\mathbf{k})$. If plane wave spreads along $z$-axis, i.e. $\mathbf{k} = \mathbf{z}_0 k_z$, then the equation (3) gives two solutions $k_z^2 = k_0^2 \varepsilon\mu$ and $k_z = \pm k_0 \sqrt{\varepsilon\mu}$, that corresponds to forward and backward waves, in which connection one may take the arithmetic value for the root, i.e. if $\varepsilon < 0$ and $\mu < 0$ then $n = \sqrt{\varepsilon\mu} > 0$. As it will be shown further, the values $\varepsilon < 0$ and $\mu < 0$ is the exactly unreliazable abstraction. So, the choice of forward or backward waves is determined by the sign of $k_z$, but not of $n$. This sign in general case of dissipative media must be chosen from the condition $\mathrm{Im}(k_z) < 0$ [8], i.e. the wave with the dependence $\exp(i\omega t - ik_z z)$ must damp in media along the direction $z$ of energy transfer. This direction in dissipative media must be determined by the Pointing vector direction [8,31,32], but not by the group velocity vector (as it is made in the majority of works). Such root choice gives the backward wave if $\varepsilon' < 0$ and $\mu' < 0$: $\mathrm{Re}(k_z) < 0$. Both considerations: the present form and the form from [1] are equivalent in isotropic case, but everyone should have in view and remember that the initial for DE is the dependence $\mathbf{k} = \mathbf{k}(k_0)$, but not the $n = n(k_0)$.

All real known LHM are bianisotropic with periodic metallic inclusions of complicated form (usually these elements are pins and split ring resonators, or $\Omega$- elements, spirals, helixes and some similar configurations). The electrophysical (electromagnetic) parameters of metamaterials must be obtained by homogenization [6,33–48]. It is fulfilled by inverse problem solutions and averaging based on full-wave analysis of periodic structure. It is necessary for this to solve many times the direct problems for dispersion and field determination using the rigorous methods (for example, integral equation method, or plane wave expansion method) [48]. The homogenization is also based on the models of media, for example, in the form [45–47]

$$\mathbf{P}^e = \varepsilon_0 (\hat{\varepsilon} - \hat{I})\overline{\mathbf{E}} + c^{-1}\hat{\varsigma}\overline{\mathbf{H}} = \varepsilon_0 \left[ (\hat{\varepsilon} - \hat{I})\overline{\mathbf{E}} + Z_0 \hat{\varsigma}\overline{\mathbf{H}} \right],$$

$$\mathbf{P}^m = \mu_0 (\hat{\mu} - \hat{I})\overline{\mathbf{H}} + c^{-1}\hat{\varsigma}\overline{\mathbf{E}} = \mu_0 \left[ (\hat{\mu} - \hat{I})\overline{\mathbf{H}} + Z_0^{-1}\hat{\varsigma}\overline{\mathbf{E}} \right],$$

and then on the determination of parameters of such models by strict or approximate fitting to the boundary problem solution [45,47]. Here $Z_0 = \sqrt{\mu_0/\varepsilon_0}$ is the vacuum impedance, $\mathbf{P}^e$ and $\mathbf{P}^m$ are averaged over the periodic cell dipole moments (electric and magnetic), the upper line means the field averaging. The higher averaged multipole moments in principle also must be included in polarization. The effective medium tensors $\hat{\varepsilon}, \hat{\mu}, \hat{\varsigma}, \hat{\xi}$ are resulted from the homogenization are dependent on averaging method and determined, at least, for wave length $\lambda > D$, where $D$ is the characteristic dimension connected with the region of averaging (for example, the cell period). The homogenization procedure in addition to calculation of averaged cell dipole moments may be based on least-square analysis (minimization) of rigorous (full-wave) and model DE solutions, or least-square analysis (minimization) for corresponding plane wave diffraction results for structure vacuum-metamaterial with different angles of hade and polarizations to boundary [10, 45, 48]. It is so as the Ewald-Oseen extinction theorem [43] in this case may be proven. One of the first such publication in which the effective permittivity was determined by plane wave normal fall on plane boundary of media with periodically included small ferrite and metallic balls, and also air-bladders (halls) in dielectric was the monograph [10]. The effective parameters in general case must be fitted so that the least-square discrepancy has the minimum [45, 48]. Let's write down the averaged fields (denoted by the upper line) as

$$\overline{\mathbf{E}} = \mathbf{A}\exp(i\omega t \mp i\mathbf{kr}), \qquad \overline{\mathbf{H}} = \mathbf{C}\exp(i\omega t \mp i\mathbf{kr}). \qquad (4)$$

In general case one can extract from Maxwell equations not the relation (1) but the following matrix equation [47]

$$\begin{bmatrix} \hat{\varepsilon} & \hat{k}/k_0 + \hat{\xi} \\ \hat{\varsigma} - \hat{k}/k_0 & \hat{\mu} \end{bmatrix} \cdot \begin{pmatrix} \mathbf{A} \\ Z_0 \mathbf{C} \end{pmatrix} = \begin{pmatrix} \mathbf{0} \\ \mathbf{0} \end{pmatrix},$$

which is equivalent to two DEs in the forms

$$\left[\left(k_0^{-1}\hat{k} + \hat{\xi}\right)\hat{\mu}^{-1}\left(k_0^{-1}\hat{k} - \hat{\varsigma}\right) + \hat{\varepsilon}\right]\mathbf{A} = 0, \quad \left[\left(k_0^{-1}\hat{k} - \hat{\varsigma}\right)\hat{\varepsilon}_e^{-1}\left(k_0^{-1}\hat{k} + \hat{\xi}\right) + \hat{\mu}_e\right]\mathbf{C} = 0 \qquad (5)$$

and to two DEs in the forms

$$\det\!\left(\left(k_0^{-1}\hat{k} + \hat{\xi}\right)\hat{\mu}^{-1}\left(k_0^{-1}\hat{k} - \hat{\varsigma}\right) + \hat{\varepsilon}\right) = 0, \quad \det\!\left(\left(k_0^{-1}\hat{k} - \hat{\varsigma}\right)\hat{\varepsilon}^{-1}\left(k_0^{-1}\hat{k} + \hat{\xi}\right) + \hat{\mu}\right) = 0. \qquad (6)$$

Here we introduce the cross-polarization tensors $\hat{\xi}, \hat{\varsigma}$ and the matrixes:

$$\hat{k} = \begin{bmatrix} 0 & -k_z & k_y \\ k_z & 0 & -k_x \\ -k_y & k_x & 0 \end{bmatrix}, \quad \hat{k}^2 = \begin{bmatrix} -k_z^2 - k_y^2 & k_x k_y & k_z k_z \\ k_x k_y & -k_z^2 - k_x^2 & k_x k_y \\ k_z k_z & k_x k_y & -k_y^2 - k_x^2 \end{bmatrix}. \qquad (7)$$

From these equations after the homogenization one should determine the dispersion relation $k_0 = f(\mathbf{k})$. The metamaterials in general are possessed of spatial dispersion, i.e. their effective spatial-depended parameters are not local and in the $\mathbf{k}$-space they are the functions of $\mathbf{k}$. Let us summarize the essence of homogenization. Many times setting the different values and directions of $\mathbf{k}$ (the wave properties differ in different directions) and determining the corresponding values of $k_0$ from boundary problem solutions and from model of DE we are calculating the polarization and fitting the material parameters so that the wave properties in inhomogeneous structures would be on average equivalent to the plane wave properties in the model homogeneous anisotropic or bianisotropic medium. Correspondingly the material equations are equivalent on average to media particles motions under the wave influence. If there are two sorts of inclusions, and the first give the input mainly into electric polarization and the second ones for the most part into magnetic one, and in which connection they have weak electromagnetic correlations, that one may neglect the cross-polarization tensors: $\hat{\xi} = \hat{\varsigma} = 0$. Then $\left[\hat{k}\hat{\mu}^{-1}\hat{k} + k_0^2\hat{\varepsilon}\right]\mathbf{A} = 0$, $\left[\hat{k}\hat{\varepsilon}^{-1}\hat{k} + k_0^2\hat{\mu}\right]\mathbf{C} = 0$. If the matrixes (7) commute with the inverse tensor $\hat{\mu}^{-1}$, that there is $\left[\hat{k}^2 + k_0^2\hat{\mu}\hat{\varepsilon}\right]\mathbf{A} = \left[\hat{k}^2 + k_0^2\hat{n}^2\right]\mathbf{A} = 0$, where

$$\hat{n} = (\hat{\mu}\hat{\varepsilon})^{1/2} = \hat{n}' - i\hat{n}'' = \left[(\hat{\mu}'\hat{\varepsilon}' - \hat{\mu}''\hat{\varepsilon}'') - i(\hat{\mu}''\hat{\varepsilon}' + \hat{\mu}'\hat{\varepsilon}'')\right]^{1/2}. \qquad (8)$$

In the small loss case we have $\hat{n} = \hat{n}'\!\left[\hat{I} - i(\hat{\mu}''\hat{\varepsilon}' + \hat{\mu}'\hat{\varepsilon}'') \cdot \hat{n}'^{-2}/2\right]$, and

$$\hat{n}' = \mathrm{Re}(\hat{n}) = (\hat{\mu}'\hat{\varepsilon}')^{1/2}, \quad \hat{n}'' = -\mathrm{Im}(\hat{n}) = (\hat{\mu}''\hat{\varepsilon}' + \hat{\mu}'\hat{\varepsilon}'') \cdot (2\hat{n}')^{-1}. \qquad (9)$$

The problem arises here how one must extract the roots from matrixes. If $\hat{\varepsilon}'$ and $\hat{\mu}'$ are the diagonal tensors with all negative (or positive) components, then $\hat{n}'$ is the positive definite matrix. But the components may have different signs. One may also to use the tensor $\hat{\tilde{n}} = (\hat{\varepsilon}\hat{\mu})^{1/2}$. The permittivity must commutate with permeability for coincidence of these two definitions. Generally we must introduce several impedances, propagation constants and several constructions like RI in the considered bianisotropic media. The first two kinds of terms are possible and necessary. But for RI it is possible and it is better not to do this, otherwise it is connected with the troubles of root extraction from matrixes. If one directly uses the Maxwell equations (but not the wave equations) the similar problems are absent, and the different impedances and propagation constants (also several kinds) are turning out in correct forms. But the RI does not quite arise instead of this. If the Cartesian axis directions coincide with the cubic periodic cell verge directions and the metallic inclusions are symmetrically located, then we have the simplifications $\hat{\varepsilon} = \varepsilon\hat{I}$, $\hat{\mu} = \mu\hat{I}$, $\varepsilon = \varepsilon' - i\varepsilon''$, $\mu = \mu' - i\mu''$, $\varepsilon'' > 0$, $\mu'' > 0$. In general case one must take the sign in (4) in such a way that the fields damp along the direction $\mathbf{n}_0 = (\mathbf{\Pi} + \mathbf{\Pi}^*)/|\mathbf{\Pi} + \mathbf{\Pi}^*|$ of energy movement. Here the $\mathbf{\Pi} = \overline{\mathbf{E}} \times \overline{\mathbf{H}}^* / 2$ is the Pointing vector. If the tensor $\hat{n}$ is diagonal and one puts $k_x = k_y = 0$, then there are two solutions: $k_z = \pm k_0 \hat{n}_x$ and $k_z = \pm k_0 \hat{n}_y$. Here the sign in dissipative media must be taken in such manner that there was the damping along the energy propagation direction. For hypothetical medium $\varepsilon = \mu = -1$ in the ideal Veselago-Pendry lens (VPL) which cannot be realized physically, one has $k_z = -k_0 n$ (the inverse backward wave), where the RI $n = \pm\sqrt{(-1)(-1)} = \pm 1$. The mentioned exotic medium $\varepsilon = \mu = -1$ (or anti-vacuum) can not be created in form of metallic PC contrary to the statement in [49] (essentially it is mentioned already in [1]). Formally it corresponds to hypothetical diluted collisionless plasma of electric and magnetic charges (monopoles) at extremely low frequency. The rarity is essential in order to neglect the collision losses and proper plasma fields which lead to gyrotropy and spatial dispersion. As some approach to this unti-vacuum one can consider the high-frequency lossless and not created at present time magnetic semiconductors (when the frequency is less than plasma frequency and the gyromagnetic resonance frequency) can serve yet. But such media must be anisotropic and gyrotropic. The listed below demands are contradictory, that causes the difficulties in such media creation even for narrow frequency band. For the electrical (denoted by index $e$) and magnetic (index $m$) polarization current densities in hypothetic media $\varepsilon = \mu = -1$ we have $\mathbf{J}_P^e = -2i\omega\varepsilon_0 \mathbf{E}$ and $\mathbf{J}_P^m = -2i\omega\mu_0 \mathbf{H}$. These currents support the wave and are in antiphase with the fields. If we apply the Pointing theorem in complex form with polarization currents as incident ones in vacuum (that is equivalent to taking into account the media), one can get the own field energy density $U_{EM}$ in the following form $U_{EM} = U_{EM}^e + U_{EM}^m = \varepsilon_0 |\mathbf{E}|^2 /4 + \mu_0 |\mathbf{H}|^2 /4$. And for stored (electric and magnetic) reactive powers in this medium one has $P_r^e = \mathbf{E}\mathbf{J}_P^{e*}/2 = i\omega\varepsilon_0 |\mathbf{E}|^2$ and $P_r^m = \mathbf{J}_P^m \mathbf{H}^*/2 = -i\omega\mu_0 |\mathbf{H}|^2$. Therefore $U_{EM}^e = U_{EM}^m$, and the reactive electric and magnetic powers are equal and counter-phased. As corresponding to this the equal averaged over the period stored electric and magnetic energy densities are the form $\langle U_{MED}^e \rangle = \langle U_{MED}^m \rangle = 2\langle U_{EM}^e \rangle = 2\langle U_{EM}^m \rangle = \langle U_{EM} \rangle$ (as in such medium $\mathbf{H} = \sqrt{\varepsilon_0/\mu_0}\mathbf{E}$). This energy is not transferred to matter, and the full energy density for field and matter is $U = 3\langle U_{EM} \rangle$. Here the brackets $\langle...\rangle$ designate the time averaging. Correspondingly the energy transport velocity in three times less than the velocity of light: $v_e = c/3$. The phase shift $\pi/2$ testifies to oscillations similar to some resonator modes. If one uses the formula (10) in [49] which is connecting $v_p$ with $v_g$ in such ideal collisionless plasma under the condition $\varepsilon = \mu = -1$, he has for phase velocity $v_p = -c$, $v_p > 0$, and $v_g = |\mathbf{v}_g| = c/3$, $\mathbf{v}_g = -\mathbf{v}_p/3$ for group velocity. Accordingly he gets $n = 1$ and the stored reactive matter energy in two time greater than the transferred by field electromagnetic energy. Here the division of energy on matter and field part is possible as there is no any interaction energy (the photon scattering is perfectly elastic). The wave movement causes the media polarization currents and they in one's turn support the wave. Just these antiphases lead to backward wave. But the energy and the majority of its carriers - photons are moving forward (from the source). It does not give the negative light pressure as the field momentum moves from the source and along the Pointing vector direction (in our case along z-axis). It must be noted that in such exotic medium and in general case of media there are always the backward and the all directional photons having the some phase shifts from the wave. The resulting collective effect describes by quasi-photons (polaritons), and the resulting energy and momentum movement goes into positive direction i.e. from the source, but the phase in case of NR runs backward as the result of interference. In this connection there is the obvious mistake in the papers [49–52]. True, it is mentioned in [52] that the pressure is positive for vacuum-LHM plane boundary and the negative pressure disappears in low frequency limit. There has been shown in the paper [53] that the Minkowski energy-momentum tensor form in the non-dispersive media is relativistic covariant, that once again testifies to Minkowski energy-momentum tensor and photon momentum in media $\mathbf{p}^M = \mathbf{D} \times \mathbf{B}$ benefit. But the introduction of RI in $\mathbf{p}^M$ for anisotropic and bianisotropic dispersive media is incompetent, including the substantiation of negative pressure and mass transferring to source when one introduces the NRI $n < 0$ (see [54]). By the way, conclusions in [49,50] contradict work [53]. According to last, if $\varepsilon$ and $\mu$ are the negative constants (from impossibility of it here we abstract), then the electromagnetic field momentum density in medium is $\mathbf{p} = \varepsilon\mu\mathbf{E}\times\mathbf{H}/c^2 = n^2\mathbf{S}/c^2$. I.e. the pressure is positive even at the negative $n$. This density is quantized value and also consists of quasiphotons momentum [55]. In [50,51] it is told about "the formula $P = \hbar k$ connecting the photon momentum value with its wave vector" (designations and citations are taken from [50]). Further the conclusion follows: "It is obvious that in case of an opposite orientation of phase and group speed when the wave vector $k$ is negative, the specified form gives negative value of an momentum of a photon, and, thereby, at absorption or reflexion of light in medium with a negative refraction index the light pressure should be replaced with a light attraction". Without concerning of slip about "a negative" vector, we will notice that, speaking about a negative direction of a vector, it is necessary to specify, concerning what. In ideal (infinite-periodic and lossless) PC all directions are equivalent, and forward (direct) and backward (return) eigenwaves are indiscernible [48]. In finite (quasi-periodic) PC (plate) there are radiating losses because of periodicity infringement. Such PC layer is the multiband filter with strong attenuation in bandgap zones and with zones of a relative transparency. Here waves are forced, and it is important, where there is a source: at the left or on the right. If it is inside of a plate it is important as concerning how the observation point is located. The Pointing vector $\mathbf{S}$ flow goes from a source, and the phase in LHM can move as from a source (a forward wave), and to it (a backward wave). If a source at the left, the photon (quasiphoton) momentum in medium has the form $\mathbf{z}_0|n|k_0\hbar = \mathbf{z}_0|n|\omega\hbar/c$ [54,55] without dependence how the phase moves. Such approach is used in [1] for construction of a beam picture: there the direction of a stream of energy is defined by a direction of a beam falling from a source.

Here it is appropriate to consider the question: wherefrom the backward wave arises? Let the source with carrier frequency $\omega$ has started to operate (has arose) at the instant $t_0 = 0$. In the homogeneous medium at the big time $t$ it creates only forward (direct and inverse) quasi-monochromatic waves of both directions. In the inhomogeneous (for example, in periodic) medium there are the reflections from their elements. The reflections come to source from both directions with tardiness, in which connection as the delay greater, the farther ele-ments are located. As the interference result of multiple reflections at the instant $t \to \infty$ it may be that the phase is moving to the source whereas the energy and momentum are always traveling from the source. Therefore the negative light pressure - that's impossible and misunderstanding.

The radiated in both directions source loses the mass (see [51]), but its momentum is zero. The mass of all closed source-field-matter system is conserved and the constant, and the lost mass is also distributed in the field (the opposite momentum phonons have the mass), and, possibly, in the medium (the losses lead to heating, and the mass of warmed-up medium is greater) [54]. The energy and momentum flows go from the source, i.e. on the right side – to the right, and on the left side – to the left.

Let us consider a problem about pressure of light in the LHM with $\varepsilon = \mu = -1$. In proposal that the constitutive (material) equations $\mathbf{D} = -\varepsilon_0 \mathbf{E}$, $\mathbf{B} = -\mu_0 \mathbf{H}$ are true for nonstationary Maxwell equations (i.e. for any frequency) then we get $\mathbf{p}^M = \mathbf{D} \times \mathbf{B} = \mathbf{E} \times \mathbf{H}/c^2$. So the pressure is such like in the vacuum and positive in accordance with the momentum direction (cf. with the reasoning in [51]). The monochromatic wave does not press on the boundary vacuum-anti-vacuum and does not transmit the momentum to such medium. Further we will show that such approach is incorrect even for monochromatic wave. Let's note that all quantities here are unambiguously defined (no any roots). But the wave pulse of train will produce the pressure, as that such medium possessed the properties inherent in it, the very large (strictly speaking, infinitely large) time is necessary for accumulation of energy of its internal (intrinsic) oscillations. For more detailed consideration let's introduce the model of rarefied plasma with electric and hypothetical magnetic charges [1]: $\varepsilon(\omega) = 1 - \omega_{pe}^2 /[\omega(\omega - j\omega_{ce})]$, $\mu(\omega) = 1 - \omega_{pm}^2 /[\omega(\omega - j\omega_{cm})]$. We regard that the harmonic plane wave propagates in this media with the electric field polarization along the axis $x$ ($E_x = E$), and the magnetic field directed along the $y$-axis ($H_y = H$). As distinct from [1, 49] we have taken into account the collisions here. Further we consider $\omega >> \max(\omega_{ce}, \omega_{cm})$ and $\omega_{pe} = \omega_{pm} = \omega_p$, from which we have $\varepsilon(\omega) = 1 - \omega_p^2/\omega^2 - j\sigma_e/(\varepsilon_0 \omega)$, $\mu(\omega) = 1 - \omega_p^2/\omega^2 - j\sigma_m/(\mu_0 \omega)$. If $\omega \approx \omega_p/\sqrt{2}$, then $\varepsilon(\omega) \approx -1 - j\sigma_e/(\varepsilon_0 \omega_p)$, $\mu(\omega) = -1 - j\sigma_m/(\mu_0 \omega_p)$. Here $\sigma_e = \varepsilon_0 \omega_p^2/\omega_{ce}$, $\sigma_m = \mu_0 \omega_p^2/\omega_{cm}$. It may be seemed that this wave complies with Maxwell equations in the form $\partial_z H = \varepsilon_0 \partial_t E - \sigma_e E$, $\partial_z E = \mu_0 \partial_t H + \sigma_m H$. If we get the balance equation for momentum from these equations using the well-known way (see [54,55]), that we find $\partial_z U_0 + \partial_t g^M = -f_e^L - f_m^L$. Here the Lorentz forces acting on the charges are in the right part, and term in the left part $g^M = S/c^2$ is the momentum density, $S = EH$ is the z-component of Pointing vector. The balance equation has the standard form, but the energy density of wave $U_0 = -(\varepsilon_0 E^2 + \mu_0 H^2)/2$ is negative (cf. with the argumentations in [1]). This density (by the implication of balance equation) is the momentum flow density in $z$-direction, therefore it may seems, that the momentum is really carried back, and the wave pressure is negative. But this is not the case. The power balance in considered equation forms also leads to such negative energy density $U_0$. We have made the gross error in our considerations by introducing the constants into nonstationary equations. Here the strict and taking into account of frequency (time) dispersion consideration is necessary, though the wave is monochromatic. This was indicated also is [1,51]. Such analysis in quasi-mono-chromatic approach gives the positive energy (see [32]) and the positive pressure. Here one must use the integral relations ([25], formula 77.3) between the inductions and the fields, where the integral operator kernels $\varepsilon(t)$ and $\mu(t)$ are obtained from the Fourier-transforms of $\varepsilon(\omega)$ and $\mu(\omega)$. Particularly,

$$\varepsilon(t) = \delta(t) + \frac{\omega_p^2 \chi(t)}{\omega_{ce} - \omega_L}\left[\exp(-\omega_L t) - \exp(-\omega_{ce} t)\right].$$

Here $\chi(t)$ is the Heaviside function and the Landau damping is introduced here to remove the pole from zero point in the spectral permittivity function $\varepsilon(\omega)$. The plane wave may be presented here as $E = E_0 \cos(\omega t - \beta z)\exp(-\alpha z)$, and $H = H_0 \cos(\omega t - \beta z - \varphi)\exp(-\alpha z)$. If $\omega = \omega_{pe} = \omega_{pm}$ then we have $H_0 = \sqrt{\varepsilon_0/\mu_0} E_0$, and when the collision frequencies tend to zero, than $\beta \to k_0$ and the phase shift $\varphi$ and attenuation constant $\alpha$ also tend to zero. Correspondingly we get $\langle U \rangle = \langle U_0 \rangle = 3\langle U_{EM} \rangle$, $\langle D \rangle = \varepsilon_0(1 + \omega_p^2/\omega^2)\langle E \rangle$, $\langle B \rangle = \mu_0(1 + \omega_p^2/\omega^2)\langle H \rangle$, i.e. for the energy and momentum transfer velocities at the frequency $\omega_p/\sqrt{2}$ we have $v_e = v_m = c/3$, and the phase velocity is equal to $c$.

As the Dirac monopoles until now are not discovered yet, and the linear collisionless plasma can't be created in principle, the LHM with real and negative $\varepsilon < 0$ and $\mu < 0$ must be regarded as hypothetical. Also they do not satisfy causality principle [25,27]. From the equation (6) under the similar proposals one may extract the DEs in which there is the hermitian conjugated tensor $\hat{n}^*$. The difficulties of NRI introduction is discussed in the paper [44], and there is the suggestion in the paper [56] to hold always $n > 0$ taking the corresponding signs in the solutions of DE, in the Snell law formulas, and in others formulas. The present paper also uses this approach with that difference, that, even for isotropic media, it is would be better not to use the term $n$ quite. It is so as it isn't posses the required analytical properties, and in another cases it cannot be unambiguously introduced in general.

There are different metamaterial models in literature. One from them may be taken in form of [6] with the taking into account the excitation of excitons. Such model is convenient for natural crystals or for metamaterials with nanodimensional inclusions when the averaging over the physical infinitesimal volume does not work already, and their proper permittivities and permeabilities and surface impedances are incorrect for use. It is shown in [6] that in this case $\varepsilon(\omega)$ and especially $\mu(\omega)$ have restricted physical meaning. Thus, the model $n(\omega) < 0$ is the very crude model which does not fully correspond to NR physics. But it is pictorial and allows one to do any qualitative conclusions using the geometrical optic approximation that has determined its spreading. The next footstep – it is the model $\varepsilon(\omega) < 0$, $\mu(\omega) < 0$. It is more rational here to consider the complex value with $\varepsilon'(\omega) < 0$, $\mu'(\omega) < 0$ and $\varepsilon''(\omega) > 0$, $\mu''(\omega) > 0$. For the LHM having weak cross-polarization effects the next level model is the usage of complex tensor permittivity and permeability. And the general model is the bianisotropic PC [47]. There is the question arising here: somebody could create a material with NR and scalar terms $\varepsilon$ и and $\mu$ having the simultaneously negative its real parts? It is obvious that such PC must be 3-D periodic with cubic sells and similar elements in its nodes having central and axis symmetry. The split-ring resonators in DNM have not such symmetry. Possible approach here is to use the embedded 3-D-P cubic cells with various elements (resonators) orientations. The usage of magnetic semiconductor 3-D-P PC lower of ferromagnetic resonance frequency for getting $\mu'(\omega) < 0$ [1] demands the external magnetic field and leads to gyrotropia. Moreover, the losses in ferrites are quite high. Another and more useful approach is the creation of biisotripic (chiral and nonreciprocal) AM. They are described by the material equations

$$\mathbf{D} = \varepsilon_0 \varepsilon \mathbf{E} + c^{-1}(\chi - i\kappa)\mathbf{H}, \quad \mathbf{B} = \mu_0 \mu \mathbf{H} + c^{-1}(\chi + i\kappa)\mathbf{E} \qquad (10)$$

with four scalar values: permittivity, permeability, chirality $\kappa$ and nonreciprocity $\chi$. Therefore one couldn't use only $n$ and $\rho$ here [57]. We get the chiral AM if $\chi = 0$ [58], and $\kappa$ may be of both signs. For example, the chaotic implantation of ideally conductive microhelixes in transparent dielectric background may serve as chiral media [58]. The sign of $\kappa$ depends from helix winding. As far back as in 1823 Fresnel has introduced for optically active media two (but not one) RIs: for right-polarized and left-polarized waves correspondingly $n_R$ and $n_L$ with specific rotation $\pi(n_R - n_L)/\lambda$ [58]. If the right-winding and left-winding helixes are chaotically located and equiprobable, that one can try to create the medium with $\kappa = 0$ and zero specific rotation. The problem here is to get the NR in such metamaterials. Here the RI is complex, and the real losses under the NR are quite high from behind the resonances. The losses are increasing with the increase of frequency and the decrease of dimensions. Let's note that it is not necessarily to have $\varepsilon'(\omega) < 0$ and $\mu'(\omega) < 0$ for NR [9]. It is only necessarily to have the obtuse angle between $\mathbf{v}_e$ and $\mathbf{v}_p$. So the NRI is the big misunderstanding.

*Let us summarize the conclusions.* There is no real scalar RI in the NR media. It corresponds only to isotropic lossless and nondispersive media models. The usage of such RI is the very simplified models will lead to some mistakes. Both terms $\hat{n}$ and $\hat{n}^*$ which in customary meaning may correspond to $n$ are complex and tensor (for anisotropic case). In general bianisotropic case even two tensors $\hat{n}$ and $\hat{n}^*$ do not describe the LHM, and in addition one should use four complex tensors. For hypothetical case of negative terms $\varepsilon < 0$ and $\mu < 0$ we may introduce one real positive RI $n = \sqrt{\varepsilon\mu}$, taking the sign of $k_z$, corresponding to backward wave, as just $k_z$ (but not $n$) is the result of DE solution. Then this RI has the meaning of retardation $n = |k_z|/k_0 = |\mathbf{v}_p|/c$. The Fermat principle in such hypothetic media is the same as in [59] with such difference that instead of negative $n$ we use the negative light way distance as the phase moves back to energy. The Snell law is modified by change of sign [56]. The positive scalar retardation coefficient $n$ may be introduced for wave in any media and any directions $\mathbf{v}_p$ and $\mathbf{v}_e$. If the customary optic lens in the operating frequency range is absolute transparent (lossless), that the phase and group velocities are equal: $\mathbf{v}_p = \mathbf{v}_g$ [31,32]. Therefore all beams come to the lens focus in phase with equal group time of retardation $\tau_g = \tau_p = 1/\int |\mathbf{v}_p| dl = \left[c\int n dl\right]^{-1}$. But in the case of ideal VPL and focusing of normally located point dipole source all rays come to the focus in zero phase, but with different group times of retardation. These times lie in the infinite interval $8d/c \leq \tau_g < \infty$. The time delay comes to infinity for ray angle near the angle $\pi/2$ relatively of lens axis. This lens does not focus quasistationary and especially nonstationary source. Even more so it does not focus short pulse, that is particularly established in [59]. The normally located at the distance $l < d$ harmonic source must act infinitely long for focusing. In case of tangential located mono-chromatic point dipole the full focusing is absent as their fields is not azimuthally symmetric, and the dipole does not create the convergent to the focus point semispherical wave as it has been shown for normally located case [21]. Appositely, the only such simplest case of normally located dipole is considered in all papers concerning to VPL. Also let's note that for finite lens thickness $d$ and finite dipole location the Ewald-Oseen extinction theorem [43] is not proved for VLP, and, apparently, can't be proved. So the dipole located at the distance $l < d$ to the "ideally matched" VPL creates the reflected quasi-spherical wave which especially strong, than the distance $l$ smaller (it is corresponded with the microstructure influence). It is need to remember that the quasi-periodic layer of finite thickness $d$ has the radiation losses [48]. So, it is needs to solve the Maxwell equations for complicated microstructures. The rigorous wave picture of real object image must be given by the combination of 3-D vector spatial (volumetric) and/or surface spectral integral transforms from source distributions in its volume (or on its surface) over all spatial variables $k_x, k_y, k_z$ in the regions $(-\infty, \infty)$. Here all modes are included: the propagated under all angles and the damped evanescent ones. Such integral transform gives the image, i.e. transfers the source value fields from the object point $\mathbf{r}'$ to the point $\mathbf{r}$ of its observation. And the kernel of this transform is the tensor Green's function of the layer. In view of this there is always some resolution limit.

It is not necessary to consider this paper as the criticism of well-known works and articles on problem of NR. The goal here is to accent on possibility to use the electromagnetic material equations which more exactly correspond to real physical processes into the media with NR. It allows one to predict their properties more precisely including the interpretation of experimental data.